\documentclass[reprint, prd, twocolumn,floatfix]{revtex4}

\usepackage{hyperref}
\usepackage[english]{babel}
\usepackage{inputenc}
\usepackage{enumerate}
\usepackage{amsfonts}
\usepackage{amssymb}
\usepackage{amsmath}
\usepackage{dcolumn}
\usepackage{bm}
\usepackage{graphicx, graphics}
\usepackage{graphicx}
\usepackage{natbib}

\begin{document}

\preprint{APS/123-QED}

\title{Two-body light front wave functions from  general AdS/QCD models}

\author{Alfredo Vega}%
 \email{alfredo.vega@uv.cl}
\affiliation{%
 Instituto de F\'isica y Astronom\'ia, \\
 Universidad de Valpara\'iso,\\
 A. Gran Breta\~na 1111, Valpara\'iso, Chile
}

\author{Miguel Angel Martin Contreras}
\email{miguelangel.martin@uv.cl}
\affiliation{%
 Instituto de F\'isica y Astronom\'ia, \\
 Universidad de Valpara\'iso,\\
 A. Gran Breta\~na 1111, Valpara\'iso, Chile
}




\date{\today}

\begin{abstract}
In this work, we consider an extension to the matching procedure proposed by Brodsky and de Teramond to obtain the two-body wave functions in the light-front formalism for holographic models. We compute the light-front wave function (LFWF) considering different static dilaton fields and  AdS-like geometric deformations. We also prove that this procedure holds for general AdS/QCD models in asymptotically AdS geometries. 
 
\end{abstract}
\maketitle

\section{Introduction}\label{aba:sec1}


The hadronic wave function in terms of the quark and gluon degrees of freedom plays an important role doing predictions for several QCD phenomena.  But in the process to do a direct extraction of this object, some drawbacks arose. There are many non-perturbative approaches to obtain properties of the distribution amplitudes and hadronic wave functions from QCD. Some time ago,  based on the AdS/CFT correspondence, Brodsky and de Teramond suggested a matching procedure to obtain mesonic Light-Front Wave Functions (LFWF) in terms of fields dual to hadrons and electromagnetic currents on the AdS side \cite{Brodsky:2006uqa, Brodsky:2007hb}. 

The LFWF obtained from the matching procedure later was improved adding different ingredients in the QCD side. But its direct effects on the AdS side were not considered. For example, in \cite{Brodsky:2008pg} the authors develop a procedure to modify the LFWF including massive quarks, that has been used widely in the literature to calculate hadron properties (e.g., see \cite{Vega:2009zb, Ahmady:2016ujw, Bacchetta:2017vzh, Chang:2017sdl, Liu:2015jna, Swarnkar:2015osa, Vega:2015hti, Momeni:2018udf, Ahmady:2019yvo, Kaur:2019kpi, Ahmady:2019hag}). Other authors introduced changes considering the t'Hooft model \cite{Chabysheva:2012fe} or looking for the consistency with the Drell-Yan-West relation between the large momentum transferred behavior observed in the nucleon electromagnetic form factors and the large-$x$ behavior of the structure functions and with the quark counting rules, producing a LFWF with an arbitrary twist \cite{Gutsche:2013zia}.

As we mentioned above, in all of the extensions done for the holographic LFWF, authors have considered the holographic proposal with the  AdS metric and quadratic dilaton as an initial ansatz and introduced changes in the QCD side without paying attention to what happens with the holographic model.  Up to our knowledge, there are not exist papers about holographic LFWF associated with other asymptotically AdS metrics or another kind of dilaton. This specific fact motivated us to extend the light-front holography ideas to other models. This is the road we want to explore in this work.

This paper is organized as follows. In section \ref{main-ideas} we summarize the matching procedure which allows us to relate the AdS modes with a two-body bound state LFWF. In section \ref{general-soft-wall}, we study the large $Q^2$ limit in the EOM for the modes dual to photons. We further notice that, in this limit we get the same equation as in the  traditional quadratic dilaton, used for a wide variety of AdS/QCD models. This is the key ingredient that opens the door to relate the AdS modes with the two-body LFWF for several holographic models. In section \ref{examples}, we consider four AdS/QCD models and develop their associated LFWF. We add an appendix where we discuss a relation between $g(x,Q^2)$, written in eq. \eqref{MapeoFnOnda}, with the Hankel transform of $J(z,Q^2)$. And finally, in section \ref{conclusions} we present our conclusions and summarize our work.


\section{Two-body wave function in holographic models}\label{main-ideas}

In \cite{Brodsky:2006uqa, Brodsky:2007hb}, the authors showed that based on the comparison of form factors, calculated in the light-front formalism and in the AdS/QCD models, it is possible to relate bulk modes to light-front wave functions. Below we briefly discuss this matching procedure,  including a generalization that will allow us to use this formalism with general AdS/QCD models.

In the light-front formalism, the electromagnetic form factor of the pion can be written as

\begin{equation}
 \label{FactorFormaLF}
\small{ F(Q^{2}) = 2 \,\pi \int\limits^{1}_{0} dx\, \frac{1-x}{x}
\int\limits_0^\infty d\zeta\, \zeta\, J_{0}
\biggl(\zeta Q \sqrt{\frac{1-x}{x}}\biggr)\,
\widetilde{\rho}  (x, \zeta),}
\end{equation}

\noindent where $Q^2$ is the spacelike transferred momentum 
squared; $J_0$ is the Bessel function of zero order, and  $\zeta$ is a variable defined as

\begin{equation}
 \zeta = \sqrt{\frac{x}{1-x}}
\bigg| \sum^{n-1}_{j=1} x_{j} b_{j} \bigg| \,,
\end{equation}

\noindent representing the $x$-weighted transverse impact coordinate associated with the spectator system, $b_{j}$ is the internal distance between constituents and the sum is over the number of spectators. In two-body case $\zeta^{2} = x(1-x) b^{2}$.

In \eqref{FactorFormaLF}, $\widetilde{\rho}(x, \zeta)$ is the effective transverse density of partons, which in the two-body case, is given by

\begin{equation}
 \label{RHO2}
\widetilde{\rho}_{n=2}(x,\zeta) =\frac{\left|\widetilde{\psi}_{q_{1} \bar{q}_{2}}  (x, \zeta)\right|^{2}}{A^{2} (1-x)^{2}}, 
\end{equation}

\noindent where $A$ is a normalization constant.

On the other hand, in the gravity side, considering an asymptotically AdS space with a metric defined as 

\begin{equation}
 ds^{2} = e^{2 A(z)} (\eta_{\mu\nu} dx^{\mu} dx^{\nu} - dz^{2}),
\end{equation}

\noindent where $\eta_{\mu \nu}$ is Minkowski 4D spacetime metric, $z$ is the holographic coordinate, $A(z)$ defines the warp factor for an asymptotically AdS space, i.e., $e^{2 A(z \rightarrow 0))} = \frac{R^{2}}{z^{2}}$. Additionally, the model considers a dilaton field $\phi(z)$ that breaks the conformal invariance. With these ingredients, the corresponding expression for the form factors related to scalar hadrons in AdS is 

\begin{equation}
 \label{FactorFormaAdS1}
 F(Q^2) = \int\limits_0^\infty dz\, e^{3 A(z) - \phi(z)} \Psi(z) J(Q^2, z) \Psi(z), \,,
\end{equation}

\noindent where $\Psi(z)$ and $J(Q^{2},z)$ are the AdS modes dual to scalar hadrons and photons. These modes are the solutions of the bulk EOM in the Sturm Liouville form associated with each bulk field. The latter form factor can be transformed into the following expression

\begin{equation}
 \label{FactorFormaAdS}
 F(Q^2) = \int\limits_0^\infty dz\, \Phi(z) J(Q^2, z) \Phi(z) \,,
\end{equation}
where $\Phi(z)$ corresponds to the solutions of the EOM transformed into an Schr\"odinger-like form, while 
$J(Q^2, z)$ remains as the same solution used before.

In order to generalize the ideas exposed in \cite{Brodsky:2006uqa, Brodsky:2007hb}, the key step is to write the electromagnetic current as \cite{Vega:2009zb}

\begin{equation}
\label{Current}
J(Q^2, z) = \int^{1}_{0} dx\, g(Q^2,x)\,
J_{0}\biggl(\zeta Q \sqrt{\frac{1-x}{x}}\biggr) \,.
\end{equation}

This allows us, after putting $z = \zeta$ and considering $x$ variable with the same physical interpretation as in (\ref{FactorFormaLF}), to compare both form factors, allowing us to establish a matching that gives a relationship between the AdS modes and the LFWF:

\begin{equation}
\label{MapeoFnOnda}
 \left| \widetilde{\psi}_{q_1\bar q_2}(x,\zeta) \right|^{2} = A^2 \,
 x (1-x)\, g(Q^{2},x)\, \frac{|\Phi(\zeta)|^{2}}{2\pi\zeta} \,.
\end{equation}

Here the factor $A$ is constrained by the probability condition
$P_{q_1\bar q_2} = \int_0^1 dx \int d^2 b \, 
|\widetilde{\psi}_{q_1\bar q_2}(x,b)|^2 \leq 1
$ with $P_{q_1\bar q_2}$ being the
probability of finding the Fock valence state $|q_1 \bar q_2\rangle$
in the meson $M$.

Recalling that in two-body case we have $\zeta^{2} = x(1-x) b^{2}$, the relationship between AdS modes and the LFWF is written as follows

\begin{equation}
\label{MapeoFnOndaconb}
 \left| \widetilde{\psi}_{q_1\bar q_2}(x,b) \right|^{2} = A^2 \,
 \frac{\sqrt{x (1-x)}}{2 \,\pi\, b}\, g(Q^2,x)\, \left|\Phi\left(\sqrt{x (1-x)}\, b\right)\right|^{2}.
\end{equation}

In general, is not so difficult to obtain $\Phi(\zeta)$ numerically in most of the known AdS/QCD models. But obtaining $g(Q^{2},x)$ is a problem still not addressed. The expression \eqref{MapeoFnOnda} has been used in  AdS/QCD models where $g(Q^{2},x) = 1$ as in the hard wall, and in the soft wall with quadratic dilaton. In the latter, there is an extra condition: $g(Q^{2},x) = 1$, which is only achieved in the large $Q^{2}$ case. 

\subsection{Example 1: Hard-wall model}

The hard-wall model proposal considers an AdS spacetime cut by an energy scale located at $z = z_{0}$ that acts as D-brane for the bulk fields. As in the original AdS/CFT, the normalizable part of the bulk field defines the hadronic modes and the non-normalizable part is connected the Schwinger source of the hadronic operators at the conformal boundary.

In this case, the modes dual to hadrons are written as 

\begin{equation}
 \label{}
\psi(z) = C \sqrt{z} J_{L} (z \mathcal{M}),
\end{equation}

\noindent where $C$ is a normalization constant, $J_{L}(z)$ is a Bessel function, $L$ is the orbital angular momentum introduced through the relationship between the conformal dimension of modes and the scaling dimension of operators. Its eigenvalues are determined by the Dirichlet boundary conditions at $z = z_{0}$.

The EM current in this formalism is written as 

\begin{equation}
 \label{JSW}
J(Q,z) = z\,Q\,K_{1}(z Q),
\end{equation}

\noindent where $K_1(z)$ is the Bessel function of the Second kind.  

By using these expressions  in the  integral form exposed in \ref{Maestra}, we can see that $g(Q^{2}, \alpha) = g(Q^{2},x) = 1$. Therefore, we obtain the well-known integral representation

\begin{equation}
 \label{}
J(Q,z) = z\,Q\,K_{1}(z Q) = \int^{1}_{0}{dx\, 
J_{0}\biggl(\zeta \,Q \sqrt{\frac{1-x}{x}}\biggr)}.
\end{equation}

Using this equation in (\ref{MapeoFnOnda}) we obtain

\begin{equation}
 \label{}
  \widetilde{\psi}_{q_1\bar q_2}(x,\zeta)  = B \,
\sqrt{x (1-x)}\, J_{L} (\zeta \mathcal{M})\,,
\end{equation}

\noindent which is the same LFWF obtained in \cite{Brodsky:2006uqa} for the hard wall model. In this case, $\zeta$ goes from zero to a maximum value $\zeta_{0}$ and $B$ is a new constant collecting the old one and some factors appearing along calculation process.

\subsection{Example 2: Traditional soft-wall model}

Here we discuss a model that considers an AdS spacetime altogether with a quadratic dilaton $\phi(z) = \kappa^{2} z^{2}$, that breaks the conformal invariance smoothly by introducing a scale $\kappa$.

In this case, the normalizable modes are written in terms of the associated Laguerre polynomials $L_n^m(z)$ as follows:

\begin{equation}
 \label{}
 \psi(z) = A\, z^{1/2+L}\, e^{-\kappa^{2} z^{2} / 2}\, L^{L}_{n}(\kappa^{2} z^{2}),
\end{equation}
and the current is written in terms of the Tricomi $U(a,b,z)$ function:

\begin{equation}
 \label{JCompleto}
J(Q^{2},z) = \Gamma \biggr(1 + \frac{Q^{2}}{4 \kappa^{2}} \biggl) U \biggr(\frac{Q^{2}}{4 \kappa^{2}},0,\kappa^{2}z^{2} \biggl).
\end{equation}

In the appendix \ref{soft-wall-appex}, we consider the exact form of  $g(Q^{2},x)$ for this dilaton field. Here we will restrict ourselves to the large $Q^{2}$ limit discussed in \cite{Brodsky:2007hb, Vega:2009zb}, with the condition $Q^{2} \gg 4 \kappa^{2}$. At this limit it is possible to observe that

\begin{equation}
 \label{Japrox}
J(Q^{2},z) \rightarrow z Q K_{1}(z Q),
\end{equation}

\noindent the integral representation for $J(Q^{2},z)$ is the same as the one used in the hard wall case. Therefore, if we apply \eqref{MapeoFnOnda} we can obtain the following expression for the two-body wave function in the soft-wall model

\begin{equation}
\label{MapeoFnOndaEj}
 \widetilde{\psi}_{q_1\bar q_2}(x,\zeta) = B \,
\sqrt{x (1-x)}\, \zeta^{L}\, e^{-\kappa^{2} \zeta^{2}}\, L^{L}_{n}(\kappa^{2} \zeta^{2}) \,,
\end{equation}

\noindent which is the expression derived in \cite{Brodsky:2007hb}, and widely used in the literature in this version, that includes massive quarks according to the prescription discussed in \cite{Brodsky:2008pg}.

Despite the fact that we consider the large $Q^{2}$ limit to do the matching, this is not problematic since in the Fock expansion for the meson wave function the valence contribution is dominant at large $Q^{2}$ \cite{Brodsky:2007hb, Vega:2009zb}. Therefore, we can conclude that the expression \eqref{MapeoFnOndaEj} is a reasonable approach in the soft-wall model. 

\begin{center}
\begin{figure*}
  \begin{tabular}{c c}
    \includegraphics[width=3.4 in]{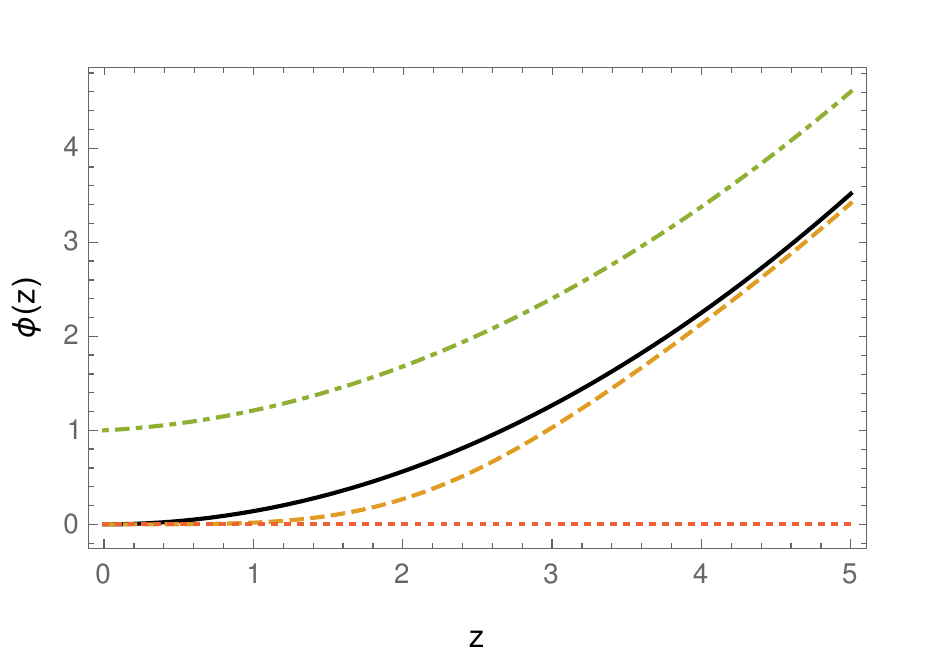}
    \includegraphics[width=3.4 in]{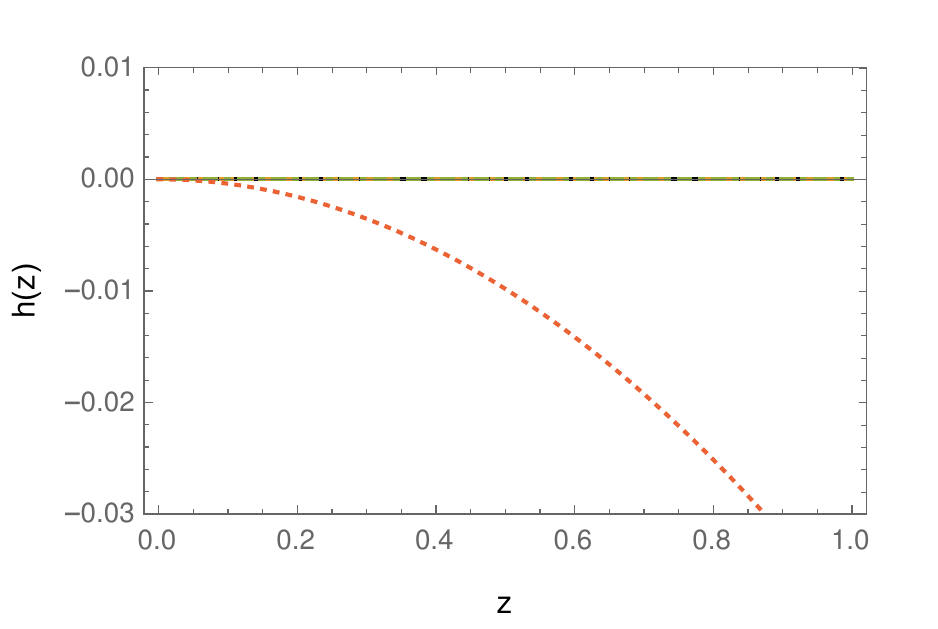}\\
    \includegraphics[width=3.4 in]{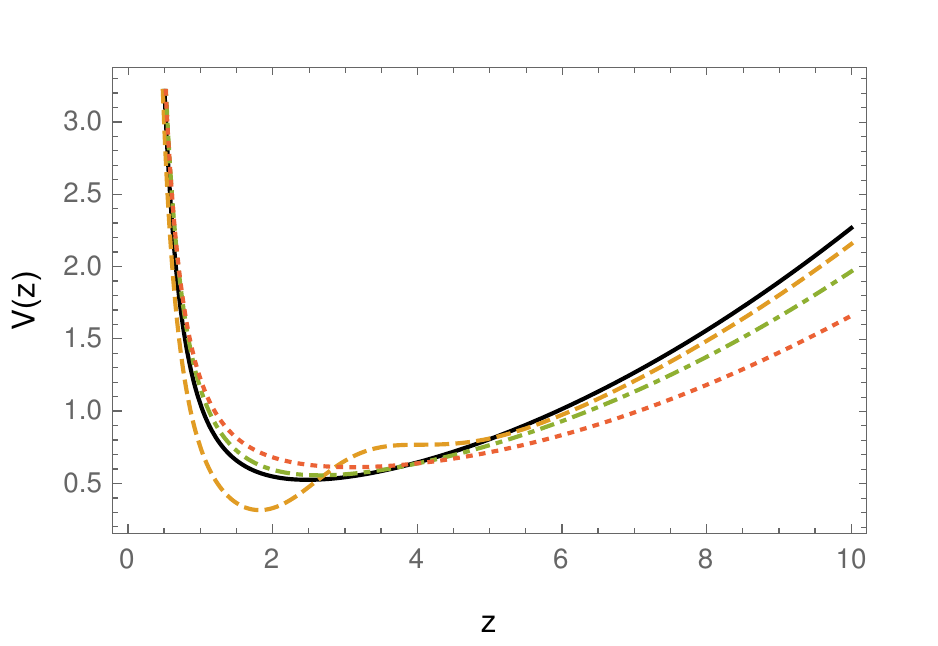}
    \includegraphics[width=3.4 in]{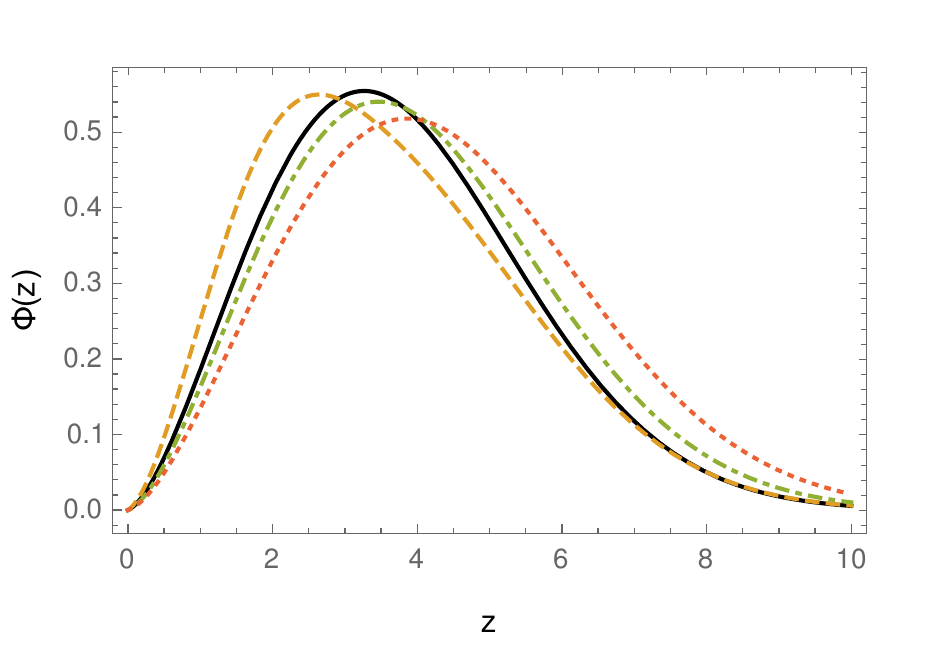}
  \end{tabular}
 \label{fig:one}
\caption{The set of dilaton fields (left upper panel), deformation function (right upper panel), holographic potentials (left lower panel) and AdS modes (right lower panel) considered in this work with the following conventions: for the model 1 we used black lines;  for model 2 we use dashed lines; for model 3  we use dot-dashed lines and model 4 is depicted with dotted lines.}
\end{figure*}
\end{center}





\section{General soft-wall model at high $Q^{2}$}\label{general-soft-wall}

As we mentioned in the previous section, the key ingredient is to put the current in the form given by the expression \eqref{Current} and then identify the $g(Q^{2},x)$ function.
As we have seen in one of the examples developed in the last section, this function is equal to one in the large $Q^{2}$ limit for the traditional soft wall model with quadratic dilaton in the AdS geometry. In this section we will analyze what happens if we consider an arbitrary dilaton and with a generic asymptotically AdS geometry. 

To do this extension, we  will consider a general AdS-like warp factor given by

\begin{equation}
\label{WarpFactor}
 A(z) = \text{ln} \left(\frac{R}{z} \right) + h(z) \,,
\end{equation}

\noindent where $R$ is the AdS radius and $h(z \rightarrow 0)$ is a deformation function that vanishes in the limit $z\rightarrow 0$.  The EOM for the current $J(Q^{2},z)$ reads as \cite{Gutsche:2011vb}

\begin{multline}\label{JGen}
\partial^{2}_{z} J(Q^{2},z)-\left[\frac{1}{z} - \partial_{z} \left(h(z) - \phi(z)\right) \right] \partial_{z} J(Q^{2},z) \\
+ Q^{2} J(Q^{2},z) = 0.
\end{multline}


Notice that to get the LFWF we consider a matching involving the expression \eqref{FactorFormaAdS}, where the modes dual to hadrons must be normalizable. Since we are interested in the current $J(Q^{2},z)$ written in the large $Q^2$ limit, holographically we can fulfill this condition with the low $z$ limit. In other words, large $Q^2$ is equivalent to the limit $z\rightarrow 0$ in this context. In this limit, the equation \eqref{JGen} reduces to 

\begin{equation}
\partial^{2}_{z} J(Q^{2},z) - \left( \frac{1}{z} \right) \partial_{z} J(Q^{2},z) + Q^{2} J(Q^{2},z) = 0 \,,
\end{equation}

\noindent which is the same equation to the hard-wall case, i.e. $g(Q^2,x)=1$. Recall that the asymptotically AdS condition is translated into the vanishing of the deformation function at the conformal boundary. This condition ensures the field/operator matching condition via the conformal dimension associated to the bulk fields \cite{Witten:1998qj}. 

Thus, for a general asymptotically AdS space we have the expression for the two-body LF wave function:

\begin{equation}
\label{MapeoFnOndaGeneral}
 \left| \widetilde{\psi}_{q_1\bar q_2}(x,\zeta) \right|^{2} = A^2 \,
 x (1-x)\, \frac{\left|\Phi(\zeta)\right|^{2}}{2\pi\zeta},
\end{equation}

\noindent that in terms of $x$ and $b$ is written as

\begin{equation}
\label{MapeoFnOndaGeneralconb}
 \left| \widetilde{\psi}_{q_1\bar q_2}(x,b) \right|^{2} = A^2 \,
 \frac{\sqrt{x (1-x)}}{2 \pi b}\, \left|\Phi\left(\sqrt{x (1-x)}\, b\right)\right|^{2},
\end{equation}
This means the expression \eqref{MapeoFnOndaconb} with $g(Q^2,x)=1$ is general and valid to obtain a two parton holographic LFWF for other models different from the usual soft-wall or hard-wall.



\section{Examples}\label{examples}
In this section we compare  the LFWF obtained for different dilatons. The recipe is as follows once we we have defined the dilaton and the AdS-like warp factor, we construct the holographic potential for scalar modes defined by 

\begin{widetext}
\begin{multline}
\label{general-pot}
V(z)=\frac{15}{4\,z^2}+\frac{1}{4}\left[\phi'(z)^2+9\,h(z)^2\right]-\frac{3}{2}\phi'(z)\,h'(z)-\frac{1}{2}\left[\phi''(z)-3\,h''(z)\right]\\
-\frac{3}{2\,z}\left[\phi'(z)-3\,h'(z)\right]+\frac{e^{2\,h(z)}M_5^2\,R^2}{z^2},
\end{multline}
\end{widetext}

\noindent where $M_5^2\,R^2$ is the bulk mass associated with the scalar modes. We will fix $M_5^2\,R^2=-3$ in our analysis. This particular choice implies that we have dual scalar meson states. 

The bulk Schrodinger-like modes are obtained by solving the Schrodinger-like EOM on the AdS side with the general holographic potential \eqref{general-pot}. With these modes, we will construct the two-body LF wave function \eqref{MapeoFnOndaGeneralconb}. 
 
\begin{center}
\begin{table*}[t]
    \begin{tabular}{||c|c|c|c|c|}
    \hline
    \textbf{Model} & \textbf{Dilaton}& \textbf{Deformation}& \textbf{Parameters} & \textbf{Ref.}\\
    \hline
    \hline
    $1$ &
    \(\displaystyle \phi_1(z)=\kappa^2\,z^2 \)
    & $h_1(z)=0$ & $\kappa=0.375$ GeV & \cite{Karch:2006pv}  \\
    \hline
    $2$ &\(\displaystyle \phi_{2}(z) = \mu_{G}^2\, z^2 \,\tanh{ \left( \frac{\mu_{G^2}^4}{\mu_{G}^2} z^2 \right)}\)& $h_2(z)=0$ & $\mu_{G} = 0.370$ GeV and $\mu_{G^2}= 0.368$ &\cite{Li:2013oda}\\
    \hline
    $3$ &\(\displaystyle \phi_3(z) = \kappa^2\, z^2 + M \,z + \,\tanh{\left( \frac{1}{M \,z} - G \right)}\) & $h_3(z)=0$& $\kappa = 0.358$ GeV, $M = 0.083$ GeV and $G = 0.082$ GeV & \cite{Braga:2019yeh}\\
    \hline
    $4$ & $\phi_4(z)=0$ & \(\displaystyle h_4(z)=\frac{1}{2}k\,z^2\) & $k=-0.280^2$ Gev$^2$ & \cite{FolcoCapossoli:2019imm}\\
    \hline
    \hline
    \end{tabular}
    \caption{Summary of the AdS/QCD models used to construct the LF wave function with their corresponding parameters and references.}
     \label{tab:one}
\end{table*}    
\end{center}

We will consider four different models from the AdS/QCD literature, characterized by their specific form of the dilaton field and the deformation function used.
A summary of these models can be found in table \ref{tab:one}. 

Notice that in the light-cone applications, the parameters involved in the wave functions are fixed by considering constraints related to the phenomenology studied. But here, in order to compare the shape of the LFWF in different cases, we need to consider a parameter set that produces the same mass for the ground state calculated from the Schrodinger-like EOM on the AdS side. A summary of the parameters used in each model is done in the table \ref{tab:one}. 

In figure 1, it is depicted a comparison between the four dilatons and deformation functions used in this paper, with their corresponding holographic potentials, given by the general expression \eqref{general-pot},  and its AdS Schrodinger-like modes calculated from such potentials.

In Figure 2 we plot the LFWF calculated for each one of the four models discussed in this paper.

\begin{center}
\begin{figure*}
  \begin{tabular}{c c}
    \includegraphics[width=3.4 in]{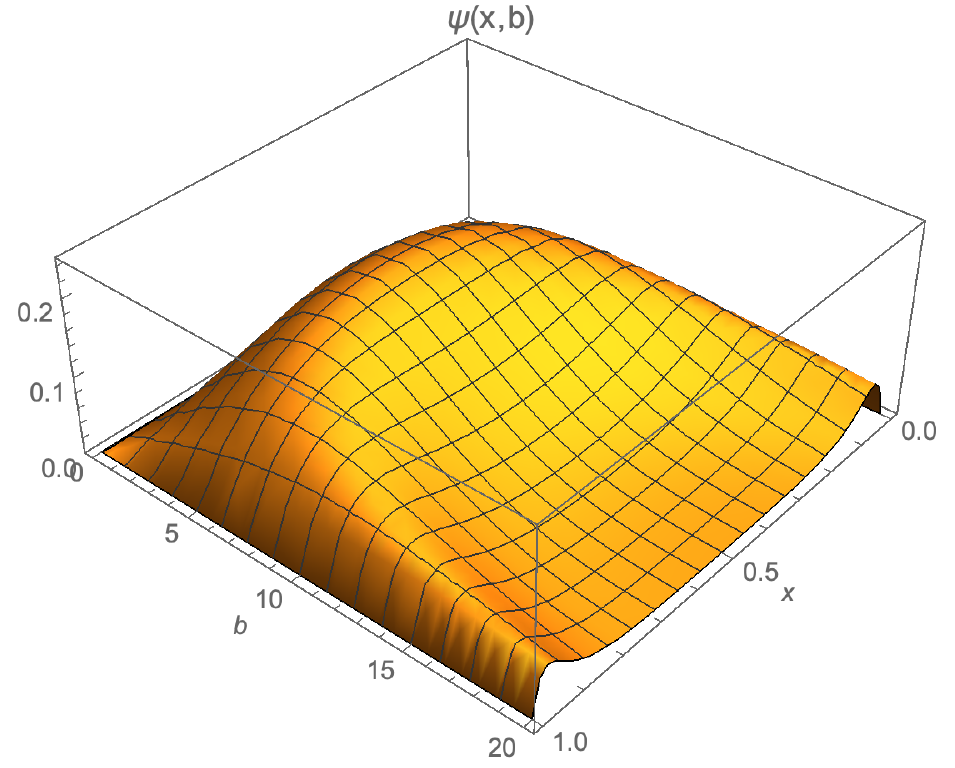}
    \includegraphics[width=3.4 in]{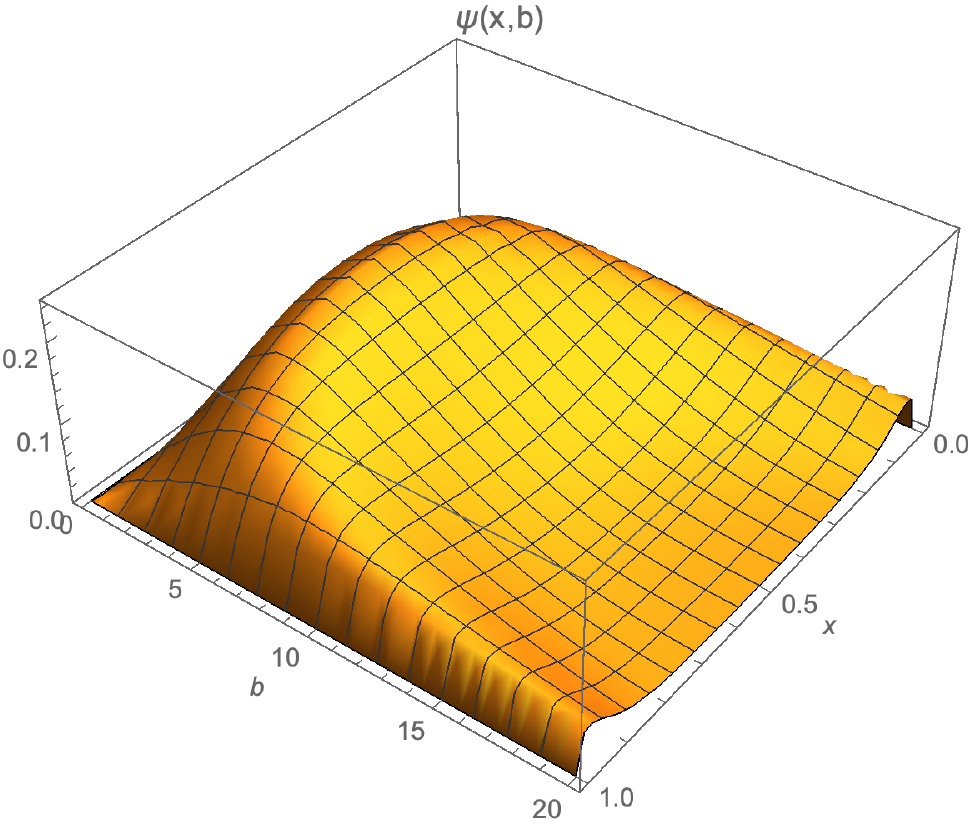}\\
    \includegraphics[width=3.4 in]{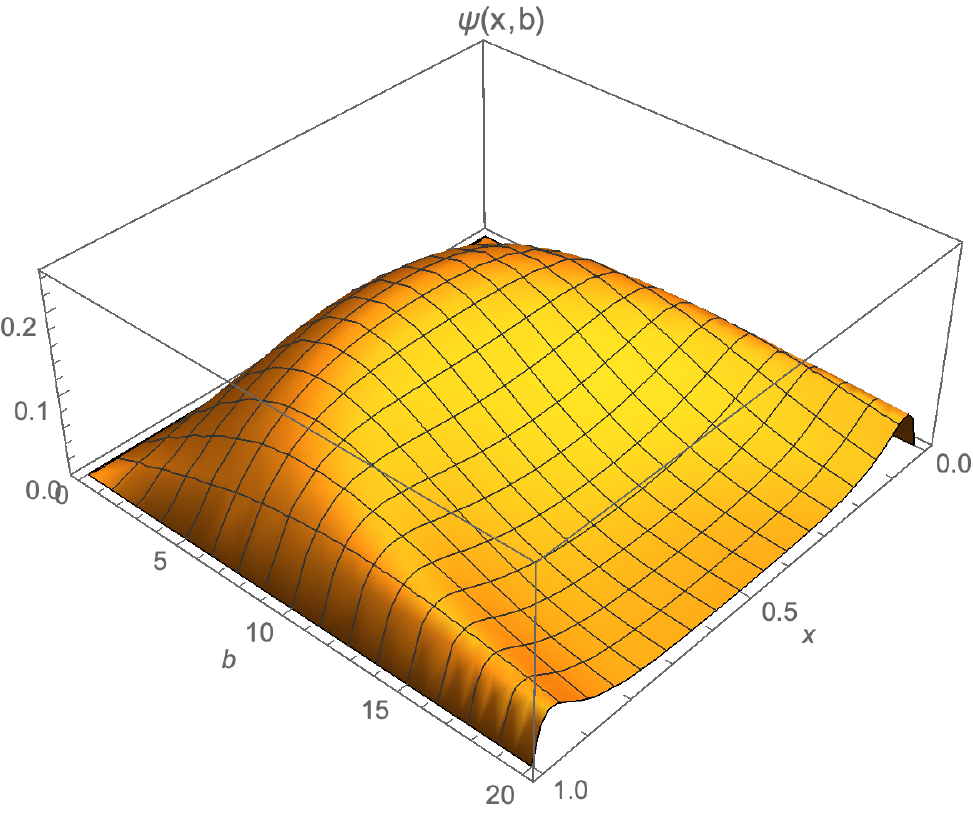}
    \includegraphics[width=3.4 in]{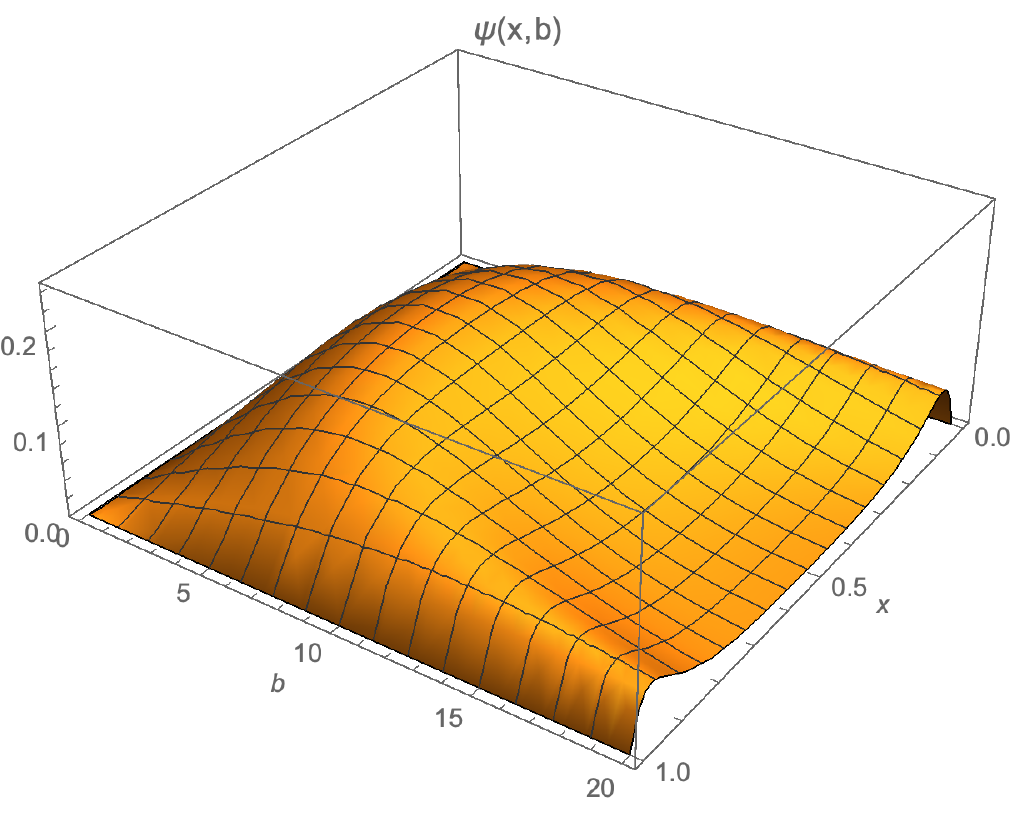}
  \end{tabular}
 \label{fig:two}
\caption{The LFWF  $\widetilde{\psi}_{q_1\bar q_2}(x,b)$ for each model discussed. Model 1 (left upper panel), model 2 (right upper panel), model 3 (left lower panel) and model 4 (right lower panel).}
\end{figure*}
\end{center}



\section{Conclusions}\label{conclusions}

The relation between AdS modes and the LFWF is an interesting topic that has been restricted the hard-wall \cite{Brodsky:2006uqa} and soft-wall models with quadratic dilaton \cite{Brodsky:2007hb} or his phenomenological modifications in the QCD side \cite{Brodsky:2008pg, Chabysheva:2012fe, Gutsche:2013zia}. 

There exists plenty of AdS/QCD models considering different dilatons, or different asymptotically AdS geometries which try to catch many of the aspects of hadronic phenomenology that the standard hard-wall or soft-wall with quadratic dilaton do not address. For these sorts of  AdS/QCD models, it was not studied their LFWF, because the matching procedure, that allows to us compare form factors at both sides and extract the LFWF in terms of AdS modes, was not discussed before in the specialized literature. Therefore, the approach considered here could be interesting, because it allows to compute the LFWF associated with these AdS/QCD approaches.

The key-point to calculate the LFWF related to AdS modes is to know properly  the distribution  $g(x,Q^2)$ defined in the expression \eqref{MapeoFnOnda}.  As it was discussed in section \ref{general-soft-wall}, for the large $Q^2$ case, this distribution is equal to one. Thus, we can obtain a two-body holographic LFWF for a wide variety of AdS/QCD models. 


In the main part of this work, we focused our attention on the simplification of the equations of motion for the vector massless field, dual to photons in the AdS side. We noticed that for a wide range of AdS/QCD models we have $g(Q^2,x)=1$. 

In the appendix, we discuss a different approach relating $g(x,Q^2)$ with the so-called Hankel transforms of the current $J(z,Q^2)$. Then we use this idea in hard-wall and soft-wall with quadratic dilaton models.

\begin{acknowledgments}
We wish to acknowledge the financial support provided by FONDECYT (Chile) under Grants No. 1180753 (A. V.) and No. 3180592 (M. A. M. C.).
\end{acknowledgments}



\appendix

\section{Extraction of $g(Q^{2},x)$ from $J(Q^{2},z)$}

In order to use the expression \eqref{MapeoFnOnda} for different AdS/QCD models, we need to compute the $g(Q^{2},x)$ function used in the current expression \eqref{Current}. To do so, we will consider the following change of variable 

\begin{equation}
 \label{alphadex}
\alpha = \sqrt{\frac{1-x}{x}}
\end{equation}

\noindent in \eqref{Current}, that will yield  the following result

\begin{equation}
 \label{}
J(Q^2, z) = \int_{0}^{\infty} d\alpha \, \frac{2 \alpha}{(1 + \alpha^{2})^{2}} \,g(\alpha) J_{0}(\zeta Q \alpha).
\end{equation}

Notice that the expression above  looks like the Hankel transform of order zero. Therefore, by using orthogonality relation 

\begin{equation}
 \label{}
\int_{0}^{\infty} J_{\nu}(kr)\, J_{\nu}(k'r) r\, dr = \frac{1}{k} \delta (k - k'),
\end{equation}

\noindent it is possible to find an expression for $g(Q^2,\alpha)$ as follows:

\begin{equation}
 \label{Maestra}
\frac{2}{Q^{2} (1 + \alpha^{2})^{2}}\, g(Q^{2},\alpha) = \int_{0}^{\infty} J(Q^{2},z) \, J_{0}(z Q \alpha) \,z \, dz.
\end{equation}

The last expression allows us to get the distribution $g(Q^{2},x)$  associated with a $J(Q^{2},z)$ for different AdS/QCD models. Let us prove this expression in the context of the hard-wall and soft-wall models. 

\subsection{Example 1: Hard Wall model}

In this case, the current has the following form

\begin{equation}
 \label{JSW}
J(Q,z) = z Q K_{1}(z Q).
\end{equation}

By using this current in integral appearing in  eqn. \eqref{Maestra} we can infer that $g(Q^{2}, \alpha) = g(Q^{2},x) = 1$,  obtaining the well-known integral representation

\begin{equation}
 \label{}
J(Q,z) = z Q K_{1}(z Q) = \int^{1}_{0} dx\, 
J_{0}\left(\zeta Q \sqrt{\frac{1-x}{x}}\right) \,.
\end{equation}

\noindent which is the  representation used in \cite{Brodsky:2006uqa} for the hard wall model.\\

\subsection{Example 2: Traditional Soft Wall model}\label{soft-wall-appex}

In this case, the current is written in terms of the Tricomi function as

\begin{equation}
 \label{JCompleto}
J(Q^{2},z) = \Gamma \left(1 + \frac{Q^{2}}{4 \kappa^{2}} \right) U \left(\frac{Q^{2}}{4 \kappa^{2}},0,\kappa^{2}z^{2} \right).
\end{equation}

In the large $Q^{2}$ limit, where $Q^{2} \gg 4 \kappa^{2}$,  it is possible to deduce that

\begin{equation}
 \label{}
J(Q^{2},z) \rightarrow z Q K_{1}(z Q).
\end{equation}

Therefore, the integral representation for $J(Q^{2},z)$ is the same as the one in the hard-wall case. But if we use the expression (\ref{JCompleto}) in the equation  (\ref{Maestra}), we obtain

\begin{widetext}
\begin{equation}
\label{}
g(Q^{2},\alpha) = \frac{\left(\alpha ^2+1\right)^2 Q^2 }{4 \kappa ^2} \left(\left(\alpha ^2+1\right) 2^{-\frac{Q^2}{2 \kappa ^2}} Q^{\frac{Q^2}{2 \kappa ^2}} \kappa ^{-\frac{Q^2}{2 \kappa ^2}} e^{\frac{\alpha ^2 Q^2}{4 \kappa ^2}} \alpha ^{2 \left(\frac{Q^2}{4 \kappa ^2}-1\right)} \Gamma \left(1-\frac{Q^2}{4 \kappa ^2},\frac{Q^2 \alpha ^2}{4 \kappa ^2}\right)-1\right),
\end{equation}

\noindent but, by using \eqref{alphadex} we have 

\begin{equation}
\label{}
g(Q^{2},x) = \frac{Q^2}{4 \kappa ^2} \left(\frac{1}{x}\right)^2 \left(\frac{2^{-\frac{Q^2}{2 \kappa ^2}} Q^{\frac{Q^2}{2 \kappa ^2}} \kappa ^{-\frac{Q^2}{2 \kappa ^2}} e^{\frac{Q^2 (1-x)}{4 \kappa ^2 x}} \left(\frac{1-x}{x}\right)^{\frac{Q^2}{4 \kappa ^2}-1} \Gamma \left(1-\frac{Q^2}{4 \kappa ^2},\frac{Q^2 (1-x)}{4 x \kappa ^2}\right)}{x}-1\right).
\end{equation}
\end{widetext}

\begin{figure}[t]
    \includegraphics[width=3.4 in]{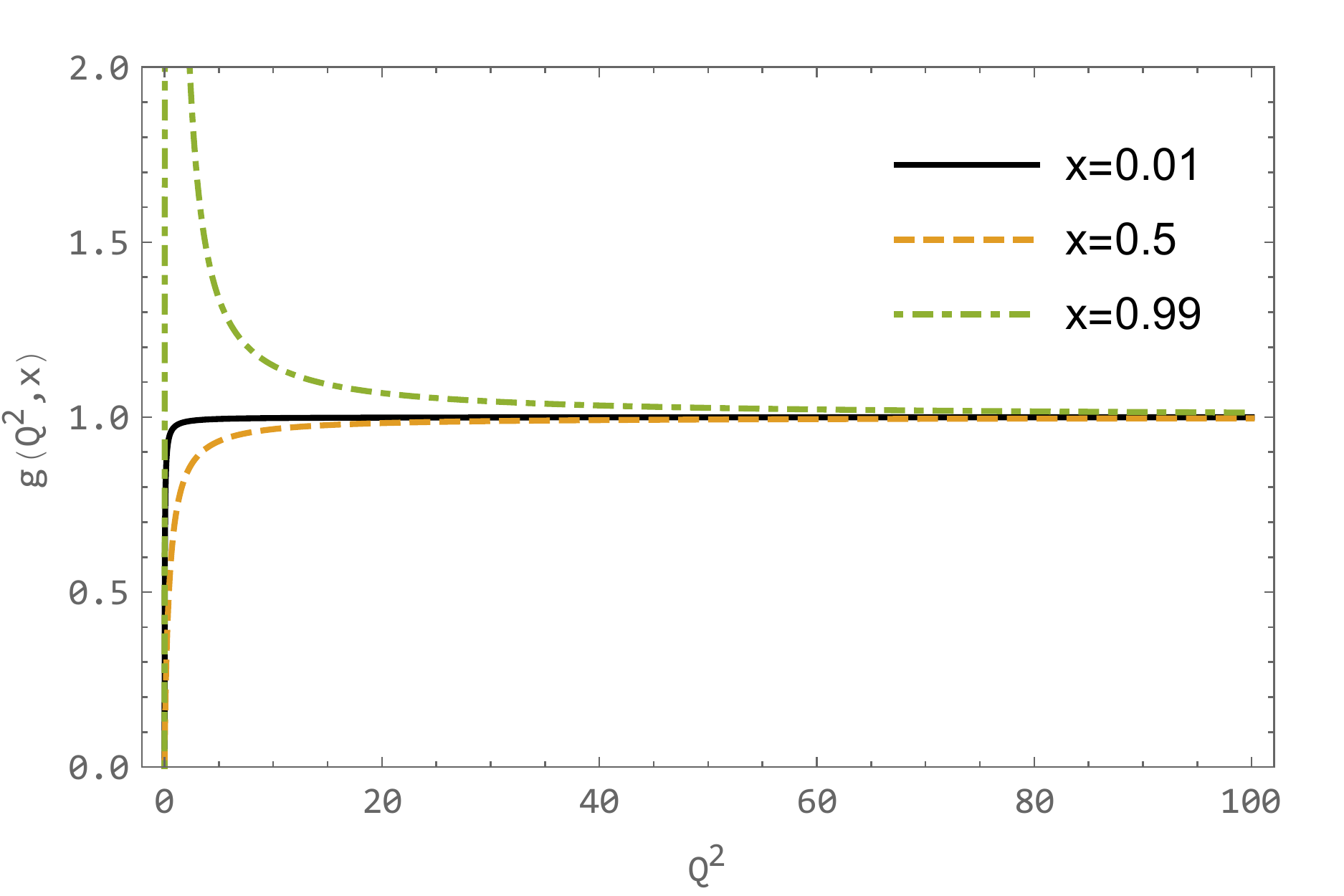}
\caption{The behavior of  $g(Q^{2},x)$ as a function of  $Q^{2}$ for different $x$ values. In all of the cases plotted, we observed $g(Q^{2} \rightarrow \infty,x) = 1$.}
\label{fig:four}
\end{figure}

As we can infer from the figure \ref{fig:four}, the distribution $g(Q^{2},x)$ goes to one at large $Q^{2}$, which is the limit where the valence contribution is dominant in the Fock expansion. Therefore, the result depicted in this work is the same presented in \cite{Brodsky:2007hb}, and used by several authors.

\bibliography{references}
\end{document}